# CP VIOLATION, MIXING, AND QUANTUM MECHANICS†

Boris Kayser
Physics Division, National Science Foundation
4201 Wilson Blvd., Arlington, VA 22230 USA

**ABSTRACT**

We discuss the quantum mechanics of B-factory experiments, and that of neutral K or B decay. Predictions for the processes to be studied at the B factories can be made through an approach based on amplitudes, rather than on wave functions. This approach avoids the puzzles of the "collapse of the wave function." In the treatment of the decay of a neutral K or B, the contributions of the different mass eigenstate components of the decaying particle must be evaluated at exactly the same spacetime point. Otherwise, the frequency predicted for the oscillation of the decay rate may be incorrect.

According to the Standard Model (SM), CP violation comes from complex phase factors in the Cabibbo-Kobayashi-Maskawa (CKM) quark mixing matrix. In the B meson system, some of the anticipated CP-violating asymmetries can cleanly probe the phases of various products of CKM elements. Thus, these asymmetries can incisively test whether CKM phases are indeed the origin of CP violation.

Most of the clean information on CKM phases will come from the decays of neutral B mesons. The neutral B meson $B_d$, a $\bar{b}d$ bound state, and its antiparticle $\overline{B_d}$, a $b\bar{d}$ bound state, mix. As a result, the $B_d$– $\overline{B_d}$ system has two mass eigenstates, $B_{Heavy}$ ($B_H$) and $B_{Light}$ ($B_L$), with complex masses

$$\mu_{H(L)} = m_B \stackrel{+}{(-)} \frac{\Delta m_B}{2} - i\frac{\Gamma}{2} \quad . \tag{1}$$

Here, $m_B$ is the average of the $B_H$ and $B_L$ masses, $\Delta m_B$ is their mass difference, and $\Gamma$ is the width which to a very good approximation they have in common.[1] Due to the mixing, a neutral B which at some proper time $\tau = 0$ is known to be a pure $|B_d\rangle$ will not remain so. Rather, after a proper time $\tau$ it will have evolved into a state $|B_d(\tau)\rangle$, given according to the SM by

$$\begin{aligned} |B_d(\tau)\rangle = e^{-i\left(m_B - i\frac{\Gamma}{2}\right)\tau} \times \\ \times\left\{\cos\left(\frac{\Delta m_B}{2}\,\tau\right)|B_d\rangle - i\,\frac{V_{td}V_{tb}^*}{V_{td}^*V_{tb}}\sin\left(\frac{\Delta m_B}{2}\,\tau\right)|\overline{B_d}\rangle\right\} \end{aligned} \tag{2}$$

Similarly, a neutral B which at $\tau = 0$ was a pure $|\overline{B_d}\rangle$ will have evolved after proper time $\tau$ into a state $|\overline{B_d}(\tau)\rangle$ which, like $|B_d(\tau)\rangle$, is a superposition of pure $|B_d\rangle$ and pure $|\overline{B_d}\rangle$. Note from Eq. (2) that, until it decays, a $|B_d(\tau)\rangle$ oscillates back and forth between being a pure $|B_d\rangle$ and a pure $|\overline{B_d}\rangle$.

Future, complementary experiments on CP violation in B decays will be carried out at dedicated high-luminosity $e^+e^-$ colliders ("B factories") and at hadron facilities. At the B factories, the B mesons will be produced via the process

$$e^+e^- \rightarrow \Upsilon(4s) \rightarrow BB \quad . \tag{3}$$

Half the time, the B mesons yielded by this process will be neutral, and it is only this case with which we shall be concerned.

Let us consider the process (3) in the Y(4s) rest frame. Since the Y(4s) has intrinsic spin J = 1, but B mesons are spinless, the B pair from $Y(4s) \to B\bar{B}$ is in a p wave. Owing to $B_d$– $\bar{B}_d$ mixing, each member of this pair oscillates back and forth between $|B_d\rangle$ and $|\bar{B}_d\rangle$ [cf. Eq. (2)]. However, owing to their common origin in the decay $Y(4s) \to B\bar{B}$, the two B mesons in the pair are correlated. In particular, they cannot both decay to the same final state f at the same time t in the Y(4s) frame.[2] For, if they did, then just after their decay we would have two identical J = 0 bosonic systems—one from each of the B mesons—in an overall p wave. This would violate the rule that two identical bosons cannot be in an antisymmetric state.

Consider, then, the situation in which the B mesons from $Y(4s) \to B\bar{B}$ decay at *different* Y(4s)-frame times $t_1$ and $t_2 > t_1$. Suppose the first decay, at time $t_1$, yields the final state $f_1$. Then, by the previous argument, we know that at time $t_1$, the B which did not yet decay must be in a state which cannot decay to $f_1$. Thus, at time $t_1$, this surviving B must be in the state $|B_{\text{Not } f_1}\rangle$ given by

$$\left|B_{\text{Not}\, f_1}\right\rangle = \left|B_d\right\rangle\left\langle f_1|T|\bar{B}_d\right\rangle - \left|\bar{B}_d\right\rangle\left\langle f_1|T|B_d\right\rangle \quad , \tag{4}$$

since, obviously, $\langle f_1|T|B_{\text{Not } f_1}\rangle = 0$. One says that the decay of the first B from $Y(4s) \to B\bar{B}$ at time $t_1$ "collapses the $B\bar{B}$ wave function," uniquely fixing the state of the remaining B at the same time. With the state of the remaining B known, one can apply the Schrödinger equation for the $B_d$ – $\bar{B}_d$ system to evolve this state forward from time $t_1$. One can then calculate the probability for this surviving B to decay to a final state $f_2$ at time $t_2$. Putting everything together, one finds, for example, that if $f_1 = \ell^- X$, a semileptonic state, and $f_2 = \Psi K_s$, then

$$\begin{aligned} &\text{Prob [One } B \to \ell^- X \text{ at } t_1;\ \text{Other } B \to \Psi K_s \text{ at } t_2] \\ &\propto e^{-\Gamma(t_2+t_1)}\left\{1-\left(\sin\varphi_{\Psi K_s}\right)\sin\left[\Delta m_B\left(t_2-t_1\right)\right]\right\}. \end{aligned} \tag{5}$$

Here, "Prob" stands for probability, and $\varphi_{\Psi K_s}$ is the phase of a certain product of CKM elements.

The expression (5), with $t_1$ and $t_2$ times in the Y(4s) rest frame, neglects the motion of the B mesons. This is an excellent approximation, since in the Y(4s) frame each B has $v/c \cong 0.06$. However, it would be desirable to have a description of the sequence $Y(4s) \to B\bar{B} \to f_1 f_2$ which is fully consistent, not only with quantum mechanics, but also with relativity. In addition, it would be advantageous to have a description which avoids the collapse of the $B\bar{B}$ wave function. One would then avoid the enigma of how, without any interaction between the two B mesons after they are born and separate, the B which decays last can know what final state the one which decays first produces, and when it does so.

What we wish to describe is a decay chain of the form

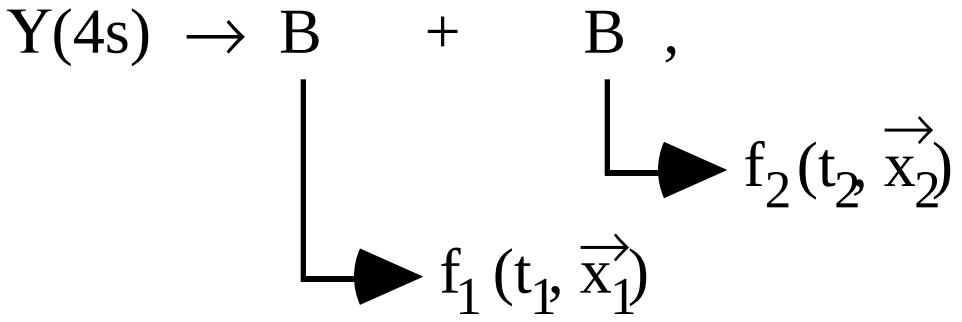

(6)

where $f_1$ is the final state into which one of the B mesons decays, $(t_1, \vec{x}_1)$ is the spacetime point at which the decay occurs, and similarly for $f_2$ and $(t_2, \vec{x}_2)$. One can describe this chain in a way which takes relativity fully into account, and avoids the collapse of the wave function, by simply calculating directly, without going through a $B\bar{B}$ wave function, the amplitude for the entire chain.[3,4] To this end, it is convenient to work in the neutral B mass eigenstate basis. The amplitude for the chain (6) has two terms. The first of these represents a process in which the B which propagates from the Y(4s) decay point to $(t_1, \vec{x}_1)$ and decays to $f_1$ is a $B_H$, while the one which propa-gates to $(t_2, \vec{x_2})$ and decays to $f_2$ is a $B_L$. The second term represents a process in which the roles of $B_H$ and $B_L$ are interchanged. Since the $B_H$–$B_L$ mass difference is tiny, these two processes are experimentally indistinguishable, so their amplitudes must be added coherently. Due to the anti-symmetry of the $\Upsilon(4s) \to BB$ amplitude, the two B mesons in (6) cannot both be $B_H$ or $B_L$.

The amplitude $A_{HL}$ for the process where the B which decays to $f_1$ ($f_2$) is a $B_H$ ($B_L$) is given by

$$A_{HL} = A(B_H \text{ to } 1; B_L \text{ to } 2)\, e^{-i\mu_H\tau_1}\, e^{-i\mu_L\tau_2} \times \\ \times\, A(B_H \to f_1)\, A(B_L \to f_2) \ . \qquad (7)$$

Here, $A(B_H$ to 1; $B_L$ to 2) is the amplitude for an $\Upsilon(4s)$ to decay to a $B_H$ moving towards $(t_1, \vec{x_1})$ and a $B_L$ moving towards $(t_2, \vec{x_2})$. (This amplitude is antisymmetric under $B_H \leftrightarrow B_L$.) The factor $\exp(-i\mu_H\tau_1)$ is the amplitude for the $B_H$ to propagate from the spacetime point where it is born to the point $(t_1, \vec{x_1})$ where it decays. In this factor, $\tau_1$ is the proper time which elapses in the $B_H$ rest frame during the propagation. Similarly, $\exp(-i\mu_L\tau_2)$ is the amplitude for the $B_L$ to propagate to $(t_2, \vec{x_2})$. [That the amplitude for a particle of mass $\mu$ to propagate for a proper time $\tau$ is $\exp(-i\mu\tau)$ follows trivially from Schrödinger's equation applied in the rest frame of the particle.] Finally, $A(B_H \to f_1)$ is the amplitude for $B_H$ to decay to $f_1$, and similarly for $A(B_L \to f_2)$. If the various A's on the right-hand side of Eq. (7) are Lorentz invariant, $A_{HL}$ is Lorentz invariant.

Adding to $A_{HL}$ the amplitude for the process in which the roles of $B_H$ and $B_L$ are interchanged, and using the antisymmetry of $A(B_H$ to 1; $B_L$ to 2), we find that the complete amplitude A for the decay chain (6) is simply

$$A \propto e^{-i\mu_H\tau_1}\, e^{-i\mu_L\tau_2}\, A(B_H \to f_1)\, A(B_L \to f_2) \\ -\, e^{-i\mu_L\tau_1}\, e^{-i\mu_H\tau_2}\, A(B_L \to f_1)\, A(B_H \to f_2) \ . \ (8)$$

If, for example, we apply this general expression to the case where $f_1 = \ell^- X$ and $f_2 = \Psi K_S$, we find that

$$\text{Prob [One } B \to \ell^- X \text{ at } \tau_1; \text{ Other } B \to \Psi K_S \text{ at } \tau_2] \\ \propto e^{-\Gamma(\tau_2+\tau_1)} \{1 - (\sin \varphi_{\Psi K_S}) \sin[\Delta m_B(\tau_2 - \tau_1)]\} \ . \ (9)$$

This result agrees with the one of Eq.(5) found by collapsing the BB wave function, except that the times $t_{1,2}$ in the rest frame of the $\Upsilon(4s)$ have been replaced by the proper times $\tau_{1,2}$ in the rest frames of the B mesons.[5] This is a negligible correction for $\Upsilon(4s) \to BB$, but a (2-3)% effect for $\varphi \to KK$, a process involving almost identical physics to be studied at the $\varphi$ factory DAΦNE.

Of course, the incorporation of small relativistic effects is not the main benefit of the amplitude approach. A more important benefit is the ability to derive theoretical expressions such as (9) which describe B-factory experiments without having to invoke the "collapse of the wave function" or to puzzle over its riddles.

---

As our discussion of B-factory experiments illustrates, the behavior of a neutral B meson involves some very interesting quantum mechanical effects. Indeed, this is true of a neutral K, D, B, or, assuming neutrino mass, a neutrino. Any of these particles is a superposition of mass eigenstates with different masses. These mass eigenstate components contribute coherently when the particle decays or interacts. Interference between these coherent contributions causes the probability for the decay or interaction to oscillate with the time or distance that the particle travels. Let us consider this oscillation.[6]

The oscillatory behavior of a particle which is a superposition of several mass eigenstates is nicely illustrated by the propagation of a kaon produced by a $K_S$ regenerator. Incident on the regenerator is a pure $K_L$ beam. With amplitude r, the regenerator introduces into the beam a $K_S$ component. Thus, a kaon emerging from the regenerator is in the state

$$|K_r\rangle = |K_L\rangle + r\,|K_S\rangle \ . \tag{10}$$

Let us call the spacetime point where the kaon leaves the regenerator (0, 0). Suppose this kaon then propagates to a spacetime point (t, x), where it is observed to decay into $\pi^+\pi^-$. Since its $|K_L\rangle$ and $|K_S\rangle$ components propagate differently, the kaon is no longer in the state $|K_r\rangle$ of Eq. (10) when it arrives at (t, x), but in a different state we shall call $|K_r(t, x)\rangle$. The probability $\Gamma[K_r(t, x) \to \pi^+\pi^-]$ for this kaon to decay to $\pi^+\pi^-$ at the point (t, x) is given by

$$\Gamma\Big[K_r(t,x) \to \pi^+\pi^-\Big] = \left| \sum_{N=S,L} A(K_r \text{ is } K_N) e^{-i\mu_N \tau^N} A(K_N \to \pi^+\pi^-) \right|^2 . \tag{11}$$

Here, $A(K_r \text{ is } K_N)$ is the amplitude for the original $K_r$ to be the mass eigenstate $K_N$, $\exp(-i\mu_N\tau^N)$ is the amplitude for this mass eigenstate to propagate from the regenerator to the decay point (t, x), and $A(K_N \to \pi^+\pi^-)$ is the amplitude for $K_N$ to decay to $\pi^+\pi^-$. In the propagation amplitude exp $(-i\mu_N\tau^N)$, $\mu_N \equiv m_N - i\Gamma_N/2$ is the complex mass of $K_N$, and $\tau^N$ is the proper time which elapses in the $K_N$ rest frame during its propagation. By including a superscript N on $\tau^N$, we are allowing for the possibility that $\tau^N$ may depend on whether the $K_N$ is a $K_S$ or $K_L$.

The precise meaning of the proper time $\tau^N$ is somewhat subtle. Since understanding its meaning is crucial to the correct treatment of particles which are mixtures of mass eigenstates, let us try to clarify what $\tau^N$ is.[3,6]

The key point is that the interfering $K_S$ and $K_L$ contributions to the decay of our kaon, like the interfering components of any spacetime-dependent wave, must be evaluated at precisely the same spacetime point (t, x).[3,6] To be sure, for a given momentum p, the $K_S$ and $K_L$ components of the kaon move at different speeds, because they have different masses. Thus, *classically* , these components, which were both born at the spacetime point (0, 0), cannot both arrive at the decay point x at the same time t. However, *quantum mechanically,* the propagating kaon is described by a wave packet, with some central momentum p. This wave packet has $K_S$ and $K_L$ pieces. Since the $K_S$ is slightly lighter than the $K_L$, the center of the $K_S$ piece of the wave packet moves faster than that of the $K_L$ piece. Thus, the centers of the $K_S$ and $K_L$ pieces become displaced relative to each other. Nevertheless, at the time t that the kaon decays, the $K_S$ and $K_L$ pieces of its wave packet overlap. It is the contributions from these two overlapping pieces at the common point x where the kaon decays that are to be added coherently. Thus, in Eq. (11), $\tau^N$ is simply the time in the $K_N$ rest frame of the decay event (t, x). That is,

$$\tau^N = \frac{1}{m_N}\big[E_N(p)t - px\big] \ , \tag{12}$$

where p is the central momentum of the wave packet, and $E_N(p) = (p^2 + m_N{}^2)^{1/2}$.

From Eq. (12), it is easily shown that through first order in $\Delta m_K \equiv m_L - m_S$, $\tau^S = \tau^L$. Thus, through first order in $\Delta m_K$, one may compute the relative phase between the two interfering amplitudes in Eq. (11) taking $\tau^S$ and $\tau^L$ to have a common value $\tau$. Experimentally, this $\tau$ is given in terms of the meaured distance x travelled by a kaon before decay and the measured momentum p of the kaon by

$$\tau = x\frac{m_K}{p} \; . \tag{13}$$

Here, $m_K \equiv \frac{1}{2}(m_L + m_S)$ is the average kaon mass.

[In complete analogy with the kaon proper times, the B meson proper times in Eqs. (7) and (8) are, through first order in $\Delta m_B$, independent of whether a $B_H$ or a $B_L$ is propagating. For this reason, we have not distinguished between the proper times for these two cases.]

With $\tau^S$ and $\tau^L$ taken to have a common value $\tau$ in Eq. (11), this expression yields the familiar result[7]

$$\begin{aligned}\Gamma[K_r(t,x)\to\pi^+\pi^-] \propto |r|^2 e^{-\Gamma_S\tau} + |\eta_{+-}|^2 e^{-\Gamma_L\tau} + \\ + 2|r||\eta_{+-}|e^{-\frac{1}{2}(\Gamma_S+\Gamma_L)\tau}\cos(\Delta m_K\tau + \varphi_r - \varphi_{+-}) \, .\end{aligned} \tag{14}$$

Here, $\eta_{+-} \equiv A(K_L\to\pi^+\pi^-) / A(K_S\to\pi^+\pi^-)$, and $\varphi_r$ and $\varphi_{+-}$ are, respectively, the phases of r and $\eta_{+-}$. The oscillatory last term on the right-hand side of Eq. (14) arises from the interference between the $K_S$ and $K_L$ contributions to the decay. Note that the frequency of oscillation of this term with $\tau$ is $\Delta m_K$. Our knowledge of $\Delta m_K$ (and similarly $\Delta m_B$) stems solely from measurements of this frequency and that of related oscillations.

What if, instead of evaluating the $K_S$ and $K_L$ contributions to the kaon decay at a common decay point x and a common decay time t, we mistakenly evaluate them at the (unequal!) $K_S$ and $K_L$ classical arrival times at the point x? The classical arrival time of $K_N$ at x is $t^N = x[E_N(p)/p]$, and the corresponding proper time is $\tau^N = x(m_N/p)$. If we use classical arrival times, then in Eq. (11) this value of $\tau^N$ replaces the quantity $\tau$ of Eq. (13). The relative phase of the $K_S$ and $K_L$ terms in Eq. (11) is then

$$m_L\tau^L - m_S\tau^S = (m_L^2 - m_S^2)\frac{x}{p} = 2(\Delta m_K)\tau \, , \tag{15}$$

where $\tau$ is the proper time of the decay determined experimentally via Eq. (13). But then the $K_S$ – $K_L$ interference term in the predicted decay rate oscillates with $\tau$ at a frequency of $2(\Delta m_K)$, rather than the correct frequency, $\Delta m_K$. That is, evaluating the contributions of the different mass eigenstates at their classical times of arrival at the decay point yields an oscillation frequency double the true value.

In summary, the contributions of the different mass eigenstate components of a neutral K, D, or B to the decay of this particle must be evaluated at exactly the same spacetime point. Evaluating them this way shows that the frequency with which the decay probability oscillates is always $\Delta m$, the difference between the masses of the mass eigenstates. This is true both for an isolated neutral K, D, or B, and for one which is part of a pair created in $\phi \to$ KK or Y(4s) $\to$ BB. Several recent descriptions of neutral particle propagation which differ from our own also conclude that the oscillation frequency is $\Delta m$.[8]

Nevertheless, there is one analysis which does not agree.[9] This analysis finds that, while the probability for decay of an isolated neutral K oscillates like cos $(\Delta m_K\tau)$, the probability for $\phi \to$ KK $\to f_1 f_2$, the $\phi$-factory analogue of the decay chain (6), oscillates like cos $[2(\Delta m_K)(\tau_2 - \tau_1)]$. Thus, this analysis predicts that when the world's first $\phi$ factory is turned on, the oscillation frequency will be found to be twice what we and most others expect.

Now, inspection of the treatment of Ref. 9 shows that in dealing with $\phi \to$ KK $\to f_1 f_2$,, it takes the $K_S$ and $K_L$ components of each kaon to have a common momentum, and evaluates their contributions to the kaon decay at their classical arrival times at the decay point. As we have seen, such a procedure gives an oscillation frequency of $2(\Delta m_K)$, while the correct value is $\Delta m_K$. Thus, the frequency found in Ref. 9 for $\phi \to KK \to f_1 f_2$ is not correct. In particular, the prediction of this reference that the oscillation frequency in $\phi \to KK \to f_1 f_2$ is twice that in decay of an isolated K is not correct. This becomes especially clear when one notes that the treatment of Ref. 9, applied to B mesons, would predict that the oscillation frequency in $\Upsilon(4s) \to BB \to f_1 f_2$, $\omega_2$, is twice the frequency in decay of an isolated B, $\omega_1$. This prediction is already experimentally contradicted. Experiments at LEP and the Tevatron find that[10]

$$\omega_1 = (0.458 \pm 0.020)\ \mathrm{ps}^{-1} \ , \tag{16}$$

while measurements at ARGUS and CLEO teach us that[11]

$$\omega_2 = (0.42 \pm 0.06)\ \mathrm{ps}^{-1} \ . \tag{17}$$

Clearly, $\omega_2$ is not twice $\omega_1$, but equal to $\omega_1$.

To sum up, the quantum mechanics of the B factory experiments can be simply understood, without invoking the “collapse of the wave function,” by calculating amplitudes directly. The quantum mechanics of a propagating neutral K, D, B, or neutrino requires that the contributions of the different mass eigenstate components of this particle to its decay or interaction always be evaluated at precisely the same spacetime point.